\documentclass[12pt, draftclsnofoot, onecolumn]{IEEEtran} 
 \pdfoutput=1 

\usepackage{cite}

\usepackage[pdftex]{graphicx}

\usepackage[cmex10]{amsmath}
\usepackage{amsfonts}
\usepackage{amsthm}

\usepackage{algorithmicx}
\usepackage{algpseudocode}

\usepackage[tight,footnotesize]{subfigure}

\usepackage{url}

\newtheorem{Assumption}{Assumption}
\newtheorem{Theorem}{Theorem}

\DeclareMathOperator*{\argmax}{\arg\!\max} 
 
\DeclareMathOperator{\size}{size}

\DeclareMathOperator{\ue}{ue}

\usepackage{accents}

\newcommand{\ceil}[1]{\lceil{#1}\rceil}

\usepackage{multirow}

\begin{document}
\title{Context-Aware Proactive Content Caching with Service Differentiation in Wireless Networks}

\author{Sabrina M\"uller*, Onur Atan$^\dagger$, Mihaela van der Schaar$^\dagger$, Anja Klein*\\
*Communications Engineering Lab, TU Darmstadt, Germany, \\ \{s.mueller, a.klein\}@nt.tu-darmstadt.de\\
$^\dagger$Department of Electrical Engineering, University of California Los Angeles, USA, oatan@ucla.edu, mihaela@ee.ucla.edu %
\thanks{
This work is copyrighted by the IEEE. It has been accepted for publication in IEEE Transactions on Wireless Communications, see IEEE's electronic database under DOI: 10.1109/TWC.2016.2636139.

A preliminary version of this work has been presented in part at the IEEE International Conference on Communications~(ICC),~2016~\cite{Mueller.Atan.2016}.}}

\maketitle

\begin{abstract}
Content caching in small base stations or wireless infostations is considered to be a suitable approach to improve the efficiency in wireless content delivery. 
Placing the optimal content into local caches is crucial due to storage limitations, but it requires knowledge about the content popularity distribution, which is often not available in advance.
Moreover, local content popularity is subject to fluctuations since mobile users with different interests connect to the caching entity over time.
Which content a user prefers may depend on the user's context.
In this paper, we propose a novel algorithm for context-aware proactive caching. 
The algorithm learns context-specific content popularity online by regularly observing context information of connected users, updating the cache content and observing cache hits subsequently.
We derive a sublinear regret bound, which characterizes the learning speed and proves that our algorithm converges to the optimal cache content placement strategy in terms of maximizing the number of cache hits. 
Furthermore, our algorithm supports service differentiation by allowing operators of caching entities to prioritize customer groups. 
Our numerical results confirm that our algorithm outperforms state-of-the-art algorithms in a real world data set, with an increase in the number of cache hits of at least~$14\%$.
\end{abstract}


\section{Introduction}

Wireless networks have been experiencing a steep increase in data traffic in recent years~\cite{cisco}. 
With the emergence of smart mobile devices with advanced multimedia capabilities and the trend towards high data rate applications, such as video streaming, especially mobile video traffic is expected to increase and to account for the majority of mobile data traffic within the next few years~\cite{cisco}. 
However, despite recent advances in cellular mobile radio networks, these networks cannot keep up with the massive growth of mobile data traffic~\cite{WangChenTaleb2014}. 
As already investigated for wired networks~\cite{Borst.etal2010}, \textit{content caching} is envisioned to improve the efficiency in wireless content delivery. 
This is not only due to decreasing disk storage prices, but also due to the fact that typically only a small number of very popular contents account for the majority of data traffic~\cite{Breslau.etal1999}. 

Within wireless networks, \textit{caching at the edge} has been extensively studied  \cite{ErmanGerberHajiaghayiEtAl2011, Maddah-AliNiesen2014, GolrezaeiMolischDimakisEtAl2013, ShanmugamGolrezaeiDimakisEtAl2013, Poularakis.Tassiulas2013, PoularakisIosifidisSourlasEtAl2016, BastugBennisDebbah2014, Bastug.etal2014, BastugBennisZeydanAbdelKaratepeSalihDebbah2015, Blasco.Gunduz2014a, Blasco.Gunduz2014b, Blasco.Gunduz2014c, SenguptaAmuruTandonEtAl2014, ElBamby.etal2014, Mueller.Atan.2016}.
At the radio access network level, current approaches comprise two types of \textit{wireless local caching entities}. 
The first type are \textit{macro base stations} (MBSs) and \textit{small base stations} (SBSs) that are implemented in wireless small cell networks, dispose of limited storage capacities
and are typically owned by the \textit{mobile network operator} (MNO).
 The second type are \textit{wireless infostations} with limited storage capacities that provide high bandwidth local data communication~\cite{Blasco.Gunduz2014b, Blasco.Gunduz2014c},~\cite{GoodmanBorrasMandayamEtAl1997},~\cite{IaconoRose2002}.
Wireless infostations could be installed in public or commercial areas and could use Wi-Fi for local data communication.
They could be owned by \textit{content providers} (CPs) aiming at increasing their users' quality of experience. 
Alternatively, third parties (e.g., the owner of a commercial area) could offer caching at infostations as a service to CPs or to the users~\cite{Blasco.Gunduz2014c}.
Both types of caching entities store a fraction of available popular content in a \textit{placement phase} and serve local users' requests via localized communication in a \textit{delivery phase}. 

Due to the vast amount of content available in multimedia platforms, 
not all available content can be stored in local caches. Hence, intelligent algorithms for \textit{cache content placement} are required. 
Many challenges of cache content placement concern content popularity.
Firstly, optimal cache content placement primarily depends on the content popularity distribution, however, when caching content at a particular point in time, it is unclear which content will be requested in future. 
Not even an estimate of the content popularity distribution might be at hand. 
It therefore must be computed by the caching entity itself \cite{Bastug.etal2014, BastugBennisZeydanAbdelKaratepeSalihDebbah2015, Blasco.Gunduz2014a, Blasco.Gunduz2014b, Blasco.Gunduz2014c, SenguptaAmuruTandonEtAl2014, ElBamby.etal2014, Mueller.Atan.2016}, 
which is not only legitimate from an overhead point of view, since else a periodic coordination with the global multimedia platform would be required. 
More importantly, local content popularity in a caching entity might not even replicate global content popularity as monitored by the global multimedia platform~\cite{GillArlittLiMahanti2007, ZinkSuhGuEtAl2009, BrodersenScellatoWattenhofer2012}. 
Hence, caching entities should learn local content popularity for a \textit{proactive} cache content placement.
Secondly, different content can be favored by different users. 
Consequently, local content popularity may change according to the different preferences of fluctuating mobile users in the vicinity of a caching entity. 
Therefore, proactive cache content placement should take into account the \textit{diversity in content popularity} across the local user population.
Thirdly, the users' preferences in terms of consumed content may differ based on their contexts, such as their location \cite{BrodersenScellatoWattenhofer2012}, personal characteristics (e.g., age~\cite{MaresSun2010}, gender~\cite{HoffnerLevine2005}, personality~\cite{RentfrowGoldbergZilca2011}, mood~\cite{Zillmann1988}), or their devices' characteristics~\cite{ZhouGuoChenNieZhu2014}. 
Hence, cache content placement should be \textit{context-aware} by taking into account that content popularity depends on a user's context. 
Thereby, a caching entity can learn the preferences of users with different contexts. 
Fourthly, while its typical goal is to maximize the number of cache hits, 
cache content placement should also take into account the cache operator's specific objective.
In particular, appropriate caching algorithms should be capable of incorporating business models of operators to offer \textit{service differentiation} to their customers,
e.g., by optimizing cache content according to different prioritization levels~\cite{KoLeeAmiriCalo2003, LuAbdelzaherSaxena2004}. 
For example, if users with different preferences are connected to a caching entity, the operator could prioritize certain users by caching content favored by these users. Moreover, certain CPs' content could be prioritized in caching decisions.

In this paper, we propose a novel context-aware proactive caching algorithm, which for the first time \textit{jointly} considers the above four aspects. 
Firstly, instead of assuming a priori knowledge about content popularity, which might be externally given or estimated in a separate training phase, our algorithm learns the content popularity online by observing the users' requests for cache content. 
Secondly, by explicitly allowing different content to be favored by different users, our algorithm is especially suitable for mobile scenarios, in which users with different preferences arrive at the wireless caching entity over time.
Thirdly, we explicitly model that the content popularity depends on a user's context, such as his/her personal characteristics, equipment, or external factors, and propose an algorithm for content caching that learns this context-specific content popularity. 
Using our algorithm, a caching entity can proactively cache content for the currently connected users based on what it has previously learned, instead of simply caching the files that are popular "on average", across the entire population of users.
The learned cache content placement strategy is proven to converge to the optimal cache content placement strategy which maximizes the expected number of cache hits. 
Fourthly, the algorithm allows for service differentiation by customer prioritization. 
The contributions of this paper are as follows:
\begin{itemize}
 \item We present a context-aware proactive caching algorithm based on contextual multi-armed bandit optimization. 
 Our algorithm incorporates diversity in content popularity across the user population and takes into account the dependence of users' preferences on their context. Additionally, it supports service differentiation by prioritization.
 \item We analytically bound the loss of the algorithm compared to an oracle, which assumes a priori knowledge about content popularity. 
 We derive a sublinear regret bound, which characterizes the learning speed and proves that our algorithm converges to the optimal cache content placement strategy which maximizes the expected number of cache hits. 
\item We present additional extensions of our approach, such as its combination with multicast transmissions and the incorporation of caching decisions based on user ratings.
 \item We numerically evaluate our caching algorithm based on a real world data set. A comparison shows that by exploiting context information in order to proactively cache content for currently connected users, our algorithm outperforms reference algorithms.
\end{itemize}

The remainder of the paper is organized as follows. 
Section~\ref{Sec_Related_Work} gives an overview of related works.
In Section~\ref{Sec_SystemModel}, we describe the system model, including an architecture and a formal problem formulation.
In Section~\ref{Sec_ContextAlgo}, we propose a context-aware proactive caching algorithm.
Theoretical analysis of regret and memory requirements are provided in Sections~\ref{Sec_regret}~and~\ref{Sec_complexity}, respectively. 
In Section~\ref{Sec_Extensions}, we propose some extensions of the algorithm.
Numerical results are presented in Section~\ref{Sec_NumResults}.
Section~\ref{Sec_Conclusion} concludes the paper.

\section{Related Work}\label{Sec_Related_Work}

\begin{table*}[!t]
\renewcommand{\arraystretch}{1.1}
\caption{Comparison with related work on learning-based caching with placement and delivery phase.}
\label{Table_related_works}
   \centering
      \begin{tabular}{|c|c|c|c|c|c|}
      \hline
 & \cite{Bastug.etal2014}, \cite{BastugBennisZeydanAbdelKaratepeSalihDebbah2015} & \cite{Blasco.Gunduz2014a, Blasco.Gunduz2014b, Blasco.Gunduz2014c} & \cite{SenguptaAmuruTandonEtAl2014} & \cite{ElBamby.etal2014} & This work\\
       \hline
      Model-Free & Yes & Yes & No & Yes & Yes \\
      \hline
      Online/Offline-Learning & Offline & Online & Online & Online & Online\\
      \hline
      Free of Training Phase& No & Yes & Yes & No  & Yes\\
      \hline
      Performance Guarantees & No & Yes & No & No  & Yes\\
      \hline
      Diversity in Content Popularity & No & No & No & Yes  & Yes\\
      \hline
      User Context-Aware & No & No & No & No  & Yes\\
      \hline
      Service Differentiation & No & No & No & No  & Yes\\
      \hline
      \end{tabular}
\end{table*}

Practical caching systems often use simple cache replacement algorithms that update the cache continuously during the delivery phase.
Common examples of cache replacement algorithms are Least Recently Used (LRU) or Least Frequently Used (LFU) (see~\cite{CaoIrani1997}).
While these simple algorithms do not consider future content popularity, recent work has been devoted to developing sophisticated cache replacement algorithms by learning content popularity trends~\cite{Li2016},~\cite{LiXuSchaarEtAl2016a}. 

In this paper, however, we focus on cache content placement for wireless caching problems with a placement phase and a delivery phase. 
We start by discussing related work that assumes a priori knowledge about content popularity. 
Information-theoretic gains achieved by combining caching at user devices with a coded multicast transmission in the delivery phase are calculated in~\cite{Maddah-AliNiesen2014}. 
The proposed coded caching approach is optimal up to a constant factor.
Content caching at user devices and collaborative device-to-device communication are combined in~\cite{GolrezaeiMolischDimakisEtAl2013} to increase the efficiency of content delivery.
In~\cite{ShanmugamGolrezaeiDimakisEtAl2013}, an approximation algorithm for uncoded caching among SBSs equipped with caches is given,
which minimizes the average delay experienced by users that can be connected to several SBSs simultaneously.
Building upon the same caching architecture, in~\cite{Poularakis.Tassiulas2013}, an approximation algorithm for distributed coded caching is presented for minimizing the probability that moving users have to request parts of content from the MBS instead of the SBSs.
In~\cite{PoularakisIosifidisSourlasEtAl2016}, a multicast-aware caching scheme is proposed for minimizing the energy consumption in a small cell network, in which the MBS and the SBSs can perform multicast transmissions. 
The outage probability and average content delivery rate in a network of SBSs equipped with caches are analytically calculated in \cite{BastugBennisDebbah2014}.

Next, we discuss related work on cache content placement without prior knowledge about content popularity.
A comparison of the characteristics of our proposed algorithm with related work of this type is given in Table~\ref{Table_related_works}.
Driven by a \textit{proactive caching paradigm},~\cite{Bastug.etal2014},~\cite{BastugBennisZeydanAbdelKaratepeSalihDebbah2015} propose a caching algorithm for small cell networks based on collaborative filtering. 
Fixed global content popularity is estimated using a training set and then exploited for caching decisions to maximize the average user request satisfaction ratio based on their required delivery rates. 
While their approach requires a training set of known content popularities and only learns during a training phase, our proposed algorithm does not need a training phase, but learns the content popularity online, thus also adapting to varying content popularities.
In~\cite{Blasco.Gunduz2014a}, using a multi-armed bandit algorithm, an SBS learns a fixed content popularity distribution online by refreshing its cache content and observing instantaneous demands for cached files.
In this way, cache content placement is optimized over time to maximize the traffic served by the SBS.
The authors extend their framework for a wireless infostation in~\cite{Blasco.Gunduz2014b},~\cite{Blasco.Gunduz2014c} by additionally taking into account the costs for adding files to the cache. 
Moreover, they provide theoretical sublinear regret bounds for their algorithms.
A different extension of the multi-armed bandit framework is given in~\cite{SenguptaAmuruTandonEtAl2014}, which exploits the topology of users' connections to the SBSs by incorporating coded caching.
The approach in \cite{SenguptaAmuruTandonEtAl2014} assumes a specific type of content popularity distribution.
Since in practice the type of distribution is unknown a priori, 
such an assumption is restrictive. 
In contrast, our proposed algorithm is model-free since it does not assume a specific type of content popularity distribution.
Moreover, in \cite{Blasco.Gunduz2014a, Blasco.Gunduz2014b, Blasco.Gunduz2014c, SenguptaAmuruTandonEtAl2014}, 
the optimal cache content placement strategy is learned over time based only on observations of instantaneous demands. 
In contrast, our proposed algorithm additionally takes diversity of content popularity across the user population into account and exploits users' context information.
Diversity in content popularity across the user population is for example taken into account in~\cite{ElBamby.etal2014}, but again without considering the users' contexts. 
Users are clustered into groups of similar interests by a spectral clustering algorithm based on their requests in a training phase. 
Each user group is then assigned to an SBS which learns the content popularity of its fixed user group over time. 
Hence, in \cite{ElBamby.etal2014}, each SBS learns a fixed content popularity distribution under the assumption of a stable user population, whereas our approach allows reacting to arbitrary arrivals of users preferring different content.

In summary, compared to related work on cache content placement (see Table~\ref{Table_related_works}), our proposed algorithm for the first time \textit{jointly} learns the content popularity online, 
allows for diversity in content popularity across the user population, takes into account the dependence of users' preferences on their context and supports service differentiation.
Compared to our previous work~\cite{Mueller.Atan.2016}, we now take into account context information at a single user level, instead of averaging context information over the currently connected users. 
This enables more fine-grained learning. 
Additionally, we incorporate service differentiation and present extensions, e.g., to multicast transmission and caching decisions based on user ratings.

We model the caching problem as a multi-armed bandit problem. 
Multi-armed bandit problems~\cite{Auer.etal2002} have been applied to various scenarios in wireless communications before~\cite{MaghsudiHossain2016}, such as cognitive jamming~\cite{AmuruTekinSchaarEtAl2016} or mobility management~\cite{ShenTekinSchaar2016}. 
Our algorithm is based on \textit{contextual multi-armed bandit} algorithms~\cite{Lu.etal2010, Slivkins2014, Tekin.Schaar2015, TekinZhangSchaar2014}.
The closest related work is~\cite{TekinZhangSchaar2014}, in which several learners observe a single context arrival in each time slot and select a subset of actions to maximize the sum of expected rewards.
While~\cite{TekinZhangSchaar2014} considers multiple learners, our system has only one learner -- the caching entity selecting a subset of files to cache in each time slot.
Compared to~\cite{TekinZhangSchaar2014}, we extended the algorithm in the following directions:
We allow multiple context arrivals in each time slot, and select a subset of actions which maximize the sum of expected rewards given the context arrivals. 
In the caching scenario, this translates to observing the contexts of all currently connected users and caching a subset of files which maximize the sum of expected numbers of cache hits given the users' contexts.
In addition, we enable each arriving context to be annotated with a weight, so that if different contexts arrive within the same time slot, differentiated services can be provided per context, by selecting a subset of actions which maximize the sum of expected weighted rewards. 
In the caching scenario, this enables the caching entity to prioritize certain users when selecting the cache content, by placing more weight on files that are favored by prioritized users. 
Moreover, we enable each action to be annotated with a weight, such that actions can be prioritized for selection. 
In the caching scenario, this enables the caching entity to prioritize certain files when selecting the cache content.

 \section{System Model}\label{Sec_SystemModel}

\subsection{Wireless Local Caching Entity}
\begin{figure}[!t]
\centering
\includegraphics[width=0.6\textwidth]{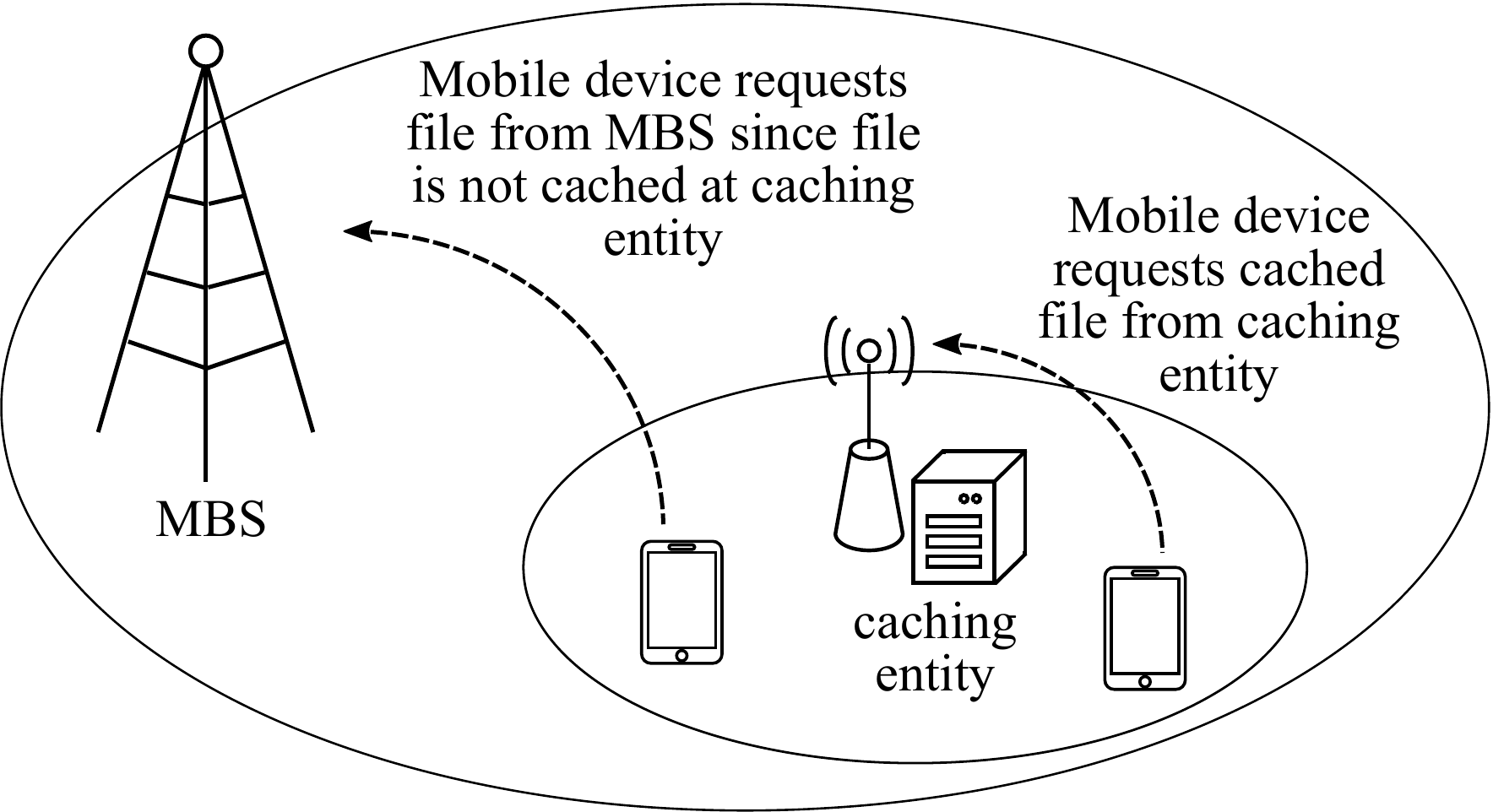}
\caption{System model.}
\label{Fig_system_model}
\end{figure}

We consider a wireless local caching entity that can either be an SBS equipped with a cache in a small cell network or a wireless infostation. 
The caching entity is characterized by a limited storage capacity and a reliable backhaul link to the core network. 
In its cache memory, the caching entity can store up to $m$ files from a finite file library $F$ containing $|F|\in \mathbb{N}$ files, where we assume for simplicity that all files are of the same size.
Users located in the coverage area can connect to the caching entity. The set of currently connected users may change dynamically over time due to the users' mobility. At most $U_{\max}\in \mathbb{N}$ users can be simultaneously connected to the caching entity.
To inform connected users about available files, the caching entity periodically broadcasts the information about the current cache content~\cite{Blasco.Gunduz2014a, Blasco.Gunduz2014b, Blasco.Gunduz2014c}.
If a user is interested in a file that the caching entity stored in its cache, the user's device requests the file from the caching entity and is served via localized communication. 
In this case, no additional load is put on neither the macro cellular network nor the backhaul network. 
If the file is not stored in the caching entity, the user's device does not request the file from the caching entity.
Instead, it requests the file from the macro cellular network by connecting to an MBS.
The MBS downloads the file from the core network via its backhaul connection, such that in this case, load is put on both the macro cellular as well as the backhaul network. 
Hence, the caching entity can only observe requests for cached files, i.e., \textit{cache hits}, but it cannot observe requests for non-cached files, i.e., \textit{cache misses}. 
Note that this restriction is specific to wireless caching and is usually not used in wired caching scenarios.
In this way, the caching entity is not congested by cache misses~\cite{Blasco.Gunduz2014a, Blasco.Gunduz2014b, Blasco.Gunduz2014c}, but learning content popularity is more difficult. 
Fig.~\ref{Fig_system_model} shows an illustration of the considered system model.

In order to reduce the load on the macro cellular network and the backhaul network, a caching entity might aim at optimizing the cache content such that the traffic it can serve is maximized, which corresponds to maximizing the number of cache hits. 
For this purpose, the caching entity should learn which files are most popular over time.

\subsection{Service Differentiation}
Maximizing the number of cache hits might be an adequate goal of cache content placement in case of an MNO operating an SBS, 
one reason being net neutrality restrictions.
However, the operator of an infostation, e.g., a CP or third party operator, 
may want to provide differentiated services to its customers (those can be both users and CPs). 
For example, if users with different preferences are connected to an infostation, the operator can prioritize certain users by caching content favored by these users. 
In this case, a cache hit by a prioritized user is associated with a higher value than a cache hit by a regular user.
For this purpose, we consider a finite set $S$ of service types. 
For service type $s \in S$, let $v_{s}\geq 1 $ denote a fixed and known weight associated with receiving one cache hit by a user of service type $s$. 
Let $v_{\max}:=\max_{s\in S} v_s$.
The weights might be selected based on a pricing policy, e.g., by paying a monthly fee, a user can buy a higher weight. 
Alternatively, the weights might be selected based on a subscription policy, e.g., subscribers might obtain priority compared to one-time users. 
Yet another prioritization might be based on the importance of users in terms of advertisement or their influence on the operator's reputation. 
Finally, prioritization could be based on usage patterns, e.g., users might indicate their degree of openness in exploring other than their most preferred content. 
Taking into account the service weights, the caching entity's goal becomes to maximize the number of \textit{weighted} cache hits.
Clearly, the above service differentiation only takes effect if users with different preferences are present, i.e., if content popularity is heterogeneous across the user population. 

Another service differentiation can be applied in case of a third party operator whose customers are different CPs. 
The operator may want to prioritize certain CPs by caching their content. 
In this case, each content is associated with a weight. 
Here, we consider a fixed and known prioritization weight $w_f\geq 1$ for each file $f\in F$ and let $w_{\max}:=\max_{f\in F} w_f$.
The prioritization weights can either be chosen individually for each file or per CP.

The case without service differentiation, where the goal is to maximize the number of (non-weighted) cache hits, 
is a special case, in which there is only one service type $s$ with weight $v_{s}=  1 $ and the prioritization weights satisfy $w_f = 1$ for all $f\in F$. 
While we refer to the more general case in the subsequent sections, this special case is naturally contained in our analysis.

\subsection{Context-Specific Content Popularity}
\begin{table}[!t]
\renewcommand{\arraystretch}{1.1}
\caption{Examples of context dimensions.}
\label{Table_context_dimensions}
   \centering
      \begin{tabular}{|c|c|}     
      \hline
      Class & Context Dimension\\
      \hline
      \multirow{2}{*}
       & demographic factors \\
      personal characteristics & personality\\
      & mood\\
      \hline
      \multirow{2}{*}
        & type of device\\
       user equipment  & device capabilities\\
       & battery status\\
      \hline
      \multirow{2}{*}
      &location\\
      external factors &time of day, day of the week\\
      & events\\
      \hline
      \end{tabular}
\end{table}
Content popularity may vary across a user population since different users may prefer different content.
A user's preferences might be linked to various factors.
We refer to such factors as \textit{context dimensions} and give some examples in Table~\ref{Table_context_dimensions}.
Relevant \textit{personal characteristics} may, for example, be demographic factors (e.g., age, gender), personality, or mood.
In addition, a user's preferences may be influenced by \textit{user equipment}, such as the type of device used to access and consume the content (e.g., smart phone, tablet), 
as well as its capabilities, or its battery status.
Besides, \textit{external factors} may have an impact on a user's preferences, such as the user's location, the time of day, the day of the week, and the taking place of events (e.g., soccer match, concert).
Clearly, this categorization is not exhaustive and the impact of each single context dimension on content popularity is unknown a priori. 
Moreover, a caching entity may only have access to some of the context dimensions, e.g., due to privacy reasons.
However, our model \textit{does not} rely on \textit{specific} context dimensions; it can use the information that \textit{is} collected from the user. 
If the caching entity does have access to some relevant context dimensions, these can be exploited to learn context-specific content popularity.

\subsection{Context-Aware Proactive Caching Architecture}
\begin{figure}[!t]
\centering
\includegraphics[width=0.6\textwidth]{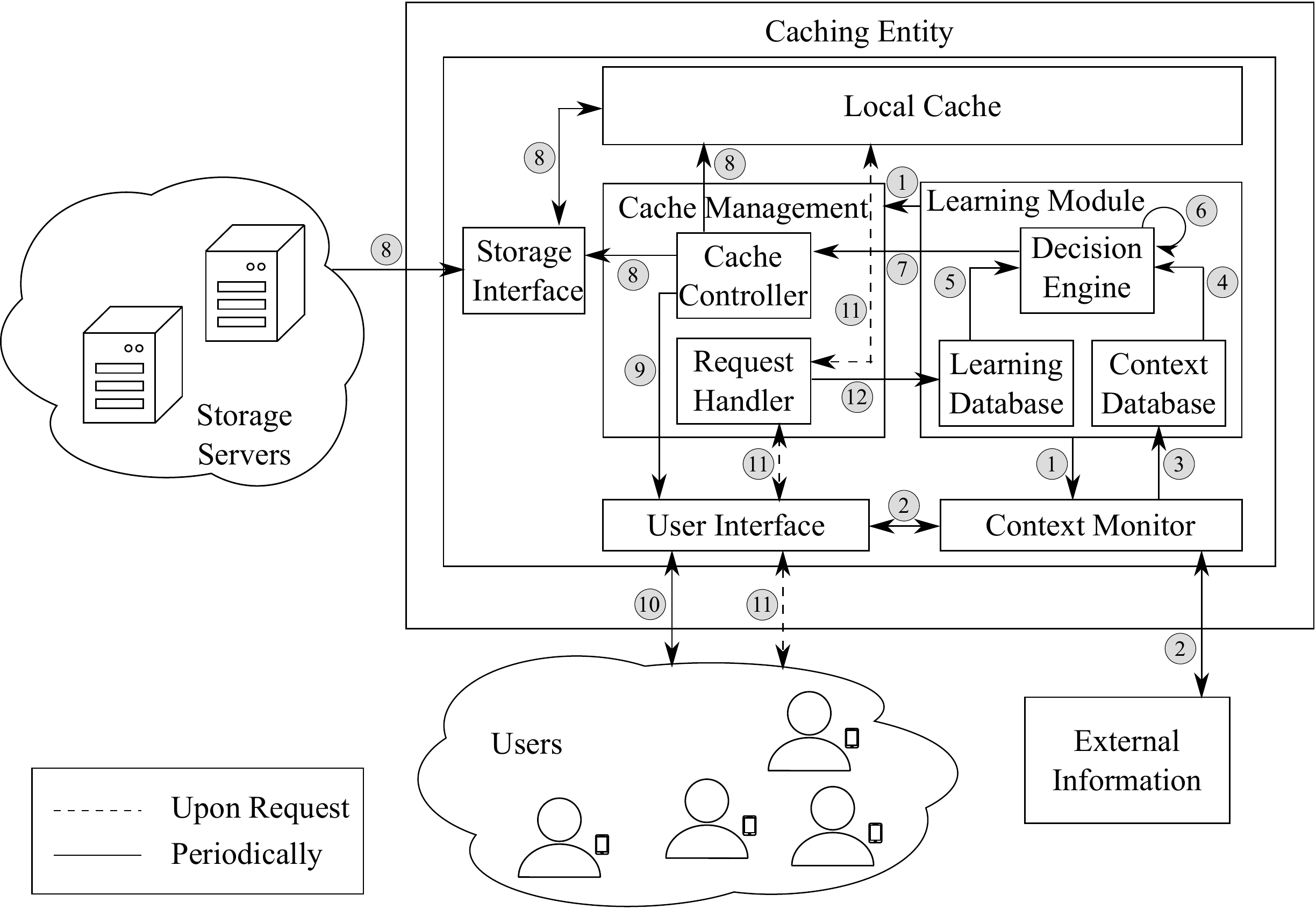}
\caption{Context-aware proactive caching architecture.}
\label{Fig_architecture}
\end{figure}

Next, we describe the architecture for context-aware proactive caching,
which is designed similarly to an architecture presented in~\cite{Li2016}.
An illustration of the context-aware proactive caching architecture is given in Fig.~\ref{Fig_architecture}.
Its main building blocks are the \textit{Local Cache}, a \textit{Cache Management} entity, a \textit{Learning Module}, a \textit{Storage Interface}, a \textit{User Interface}, and a \textit{Context Monitor}. 
The Cache Management consists of a \textit{Cache Controller} and a \textit{Request Handler}.
The Learning Module contains a \textit{Decision Engine}, a \textit{Learning Database}, and a \textit{Context Database}.
The workflow consists of several phases as enumerated in Fig.~\ref{Fig_architecture} and is described below.

\begin{itemize}
\item Initialization\\
(1)~The Learning Module is provided with the goal of caching (i.e., maximize number of cache hits or achieve operator-specific goal). 
It fixes the appropriate periodicity of context monitoring and cache refreshment.
Then, it informs the Cache Management and the Context Monitor about the periodicity.
\item Periodic Context Monitoring and Cache Refreshment\\
(2)~The Context Monitor periodically gathers context information by accessing information about currently connected users available at the User Interface and optionally by collecting additional information from external sources (e.g., social media platforms). 
If different service types exist, the Context Monitor also retrieves the service types of connected users. 
(3)~The Context Monitor delivers the gathered information to the Context Database in the Learning Module. 
(4)~The Decision Engine periodically extracts the newly monitored context information from the Context Database. 
(5)~Upon comparison with results from previous time slots as stored in the Learning Database, (6) the Decision Engine decides which files to cache in the coming time slot. 
(7)~The Decision Engine instructs the Cache Controller to refresh the cache content accordingly. 
(8)~The Cache Controller compares the current and the required cache content and removes non-required content from the cache. 
If some required content is missing, the Cache Controller directs the Storage Interface to fetch the content from storage servers and to store it into the local cache. 
(9)~Then, the Cache Controller informs the User Interface about the new cache content. 
(10)~The User Interface pushes the information about new cache content to currently connected users. 
\item User Requests\\
(11)~When a user requests a cached file, the User Interface forwards the request to the Request Handler. 
The Request Handler stores the request information, retrieves the file from the local cache and serves the request. 
\item Periodic Learning\\
(12)~Upon completion of a time slot, the Request Handler hands the information about all requests from that time slot to the Learning Module. 
The Learning Module updates the Learning Database with the context information from the beginning of the time slot and the number of requests for cached files in that time slot.
\end{itemize}

\subsection{Formal Problem Formulation}

\begin{figure}[!t]
\centering
\includegraphics[width=0.4\textwidth]{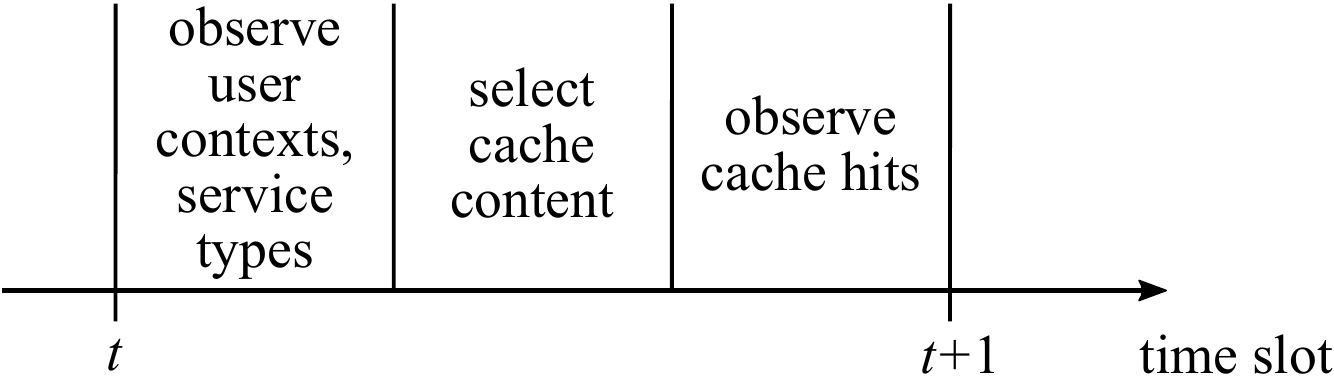}
\caption{Sequence of operations of context-aware proactive caching in time slot $t$.}
\label{Fig_algo_overview}
\end{figure}

Next, we give a formal problem formulation for context-aware proactive caching.
The caching system operates in discrete time slots $t=1,2,...,T$, where $T$ denotes the finite time horizon.
As illustrated in Fig.~\ref{Fig_algo_overview}, each time slot $t$ consists of the following sequence of operations:
(i)~The context of currently connected users and their service types are monitored. 
Let $U_t$ be the number of currently connected users.
We assume that $1\leq U_t\leq U_{\max}$ and we specifically allow the set of currently connected users to change in between the time slots of the algorithm, so that user mobility is taken into account.
Let $D$ be the number of monitored context dimensions per user. 
We denote the $D$-dimensional context space by $\mathcal{X}$. 
It is assumed to be bounded and can hence be set to $\mathcal{X}:=[0,1]^D$ without loss of generality.
Let $x_{t,i} \in \mathcal{X}$ be the context vector of user $i$ observed in time slot~$t$. 
Let $\mathbf{x_t}=(x_{t,i})_{i=1,...,U_t}$ be the collection of contexts of all users in time slot $t$.
Let $s_{t,i} \in S$ be the service type of user $i$ in time slot $t$ and let $\mathbf{s_{t}} =  (s_{t,i})_{i=1,...,U_t}$ be the collection of service types of all users in time slot~$t$.
(ii)~The cache content is refreshed based on the contexts $\mathbf{x_t}$, the service types $\mathbf{s_t}$ and their service weights, the file prioritization weights $w_f$, $f\in F$, and knowledge from previous time slots. 
Then, connected users are informed about the current cache content, which is denoted by $C_t=\{c_{t,1},...,c_{t,m}\}$.
(iii)~Until the end of the time slot, users can request currently cached files. 
Their requests are served. 
The demand $d_{c_{t,j}}(x_{t,i},t)$ of each user $i=1,...,U_t$ for each cached file $c_{t,j} \in C_t$, $j=1,...,m$, in this time slot is observed, i.e., the number of cache hits for each cached file is monitored. 

The number of times a user with context vector $x\in \mathcal{X}$ requests a file $f\in F$ within one time slot is a random variable with unknown distribution.
We denote this random demand by $d_f(x)$ and its expected value by $ \mu_f(x):=E(d_f(x)) $. 
The random demand is assumed to take values in $[0,R_{\max}]$, 
where $R_{\max}\in \mathbb{N}$ is the maximum possible number of requests a user can submit within one time slot.
This explicitly incorporates that a user may request the same file repeatedly within one time slot.
In time slot $t$, the random variables $(d_f(x_{t,i}))_{i=1,..,U_t, f\in F}$, are assumed to be independent, i.e., the requests of currently connected users and between different files are independent of each other.
Moreover, each random variable $d_f(x_{t,i})$ is assumed to be independent of past caching decisions and previous demands. 

The goal of the caching entity is to select the cache content in order to maximize the expected cumulative number of (weighted) cache hits up to the finite time horizon~$T$. 
We introduce a binary variable $y_{t,f}$, which describes if file $f$ is cached in time slot $t$, where $y_{t,f}=1$, if $f\in C_t$, and $0$ otherwise. Then, the problem of cache content placement can be formally written as
\begin{align}\label{Eq_opt_problem}
\max &\sum_{t=1}^T \sum_{f\in F} y_{t,f} w_{f} \sum_{i=1}^{U_t} v_{s_{t,i}}  \mu_{f}(x_{t,i})\\
\text{s.t.} & \sum_{f\in F}y_{t,f} \leq  m, \ \ t=1,...,T,\notag\\
&y_{t,f} \in \{0,1\}, \ \  f\in F, \ t=1,...,T. \notag
\end{align}
Let us now first assume that the caching entity \textit{had} a priori knowledge about context-specific content popularity \textit{like an omniscient oracle}, i.e., suppose that for each context vector $x\in \mathcal{X}$ and for each file $f\in F$, the caching entity would know the expected demand $\mu_f(x)=E(d_f(x))$. 
In this case, problem~\eqref{Eq_opt_problem} corresponds to an integer linear programming problem.
The problem can be decoupled into $T$ independent sub-problems, one for each time slot $t$. 
Each sub-problem is a special case of the knapsack problem \cite{KellererPferschyPisinger2004} with a knapsack of capacity~$m$ and with items of non-negative profit and unit weights. 
Hence, its optimal solution can be easily computed in a running time of $O(|F|\log(|F|))$ as follows. 
In time slot $t$, given the contexts $\mathbf{x_t}$ and the service types $\mathbf{s_t}$, 
the optimal solution is given by ranking the files according to their (weighted) expected demands and by selecting the $m$ highest ranked files.
We denote these \textit{top-$m$ files for pair~($\mathbf{x_t}$,~$\mathbf{s_t}$)} by $f_1^\ast(\mathbf{x_t},\mathbf{s_t}), f_2^\ast(\mathbf{x_t},\mathbf{s_t}),...,f_m^\ast(\mathbf{x_t},\mathbf{s_t})\in F$. 
Formally, for $j=1,...,m$, they satisfy\footnote{Several files may have the same expected demands, i.e., the optimal set of files may not be unique. This is also captured here.}
\begin{align}
f_j^\ast(\mathbf{x_t},\mathbf{s_t}) &\in \argmax _{f\in F \setminus (\bigcup_{k=1}^{j-1}\{f_k^\ast(\mathbf{x_t},\mathbf{s_t})\})} w_f \sum_{i=1}^{U_t}v_{s_{t,i}}\mu_f(x_{t,i}),
\end{align}
where $\bigcup_{k=1}^{0}\{f_k^\ast(\mathbf{x_t},\mathbf{s_t})\}:=\emptyset$.
We denote by $C_t^*(\mathbf{x_t},\mathbf{s_t}) := \bigcup_{k=1}^m\{f_k^\ast(\mathbf{x_t},\mathbf{s_t})\}$ an optimal choice of files to cache in time slot $t$.
Consequently, the collection
\begin{align}\label{Eq_opt_collection}
(C_t^*(\mathbf{x_t},\mathbf{s_t}))_{t=1,...,T}
\end{align}
is an optimal solution to problem~\eqref{Eq_opt_problem}.
Since this solution can be achieved by an omniscient oracle under a priori knowledge about content popularity, we call it the \textit{oracle solution}.

However, in this paper we assume that the caching entity \textit{does not} have a priori knowledge about content popularity. 
In this case, the caching entity cannot simply solve problem~\eqref{Eq_opt_problem} as described above, since the expected demands $\mu_f(x) = E(d_f(x))$ are unknown. 
Hence, the caching entity has to learn these expected demands over time by observing the users' demands for cached files given the users' contexts. 
For this purpose, over time, the caching entity has to find a trade-off between caching files about which little information is available (\textit{exploration}) and files of which it believes that they will yield the highest demands (\textit{exploitation}). 
In each time slot, the choice of files to be cached depends on the history of choices in the past and the corresponding observed demands. 
An algorithm which maps the history to the choices of files to cache is called a \textit{learning algorithm}.
The oracle solution given in~\eqref{Eq_opt_collection} can be used as a benchmark to evaluate the loss of learning.
Formally, the \textit{regret} of learning with respect to the oracle solution is given by
\begin{align}\label{Eq_regret}
R(T) = \sum_{t=1}^T \sum _{j=1}^m \sum _{i=1}^{U_t} v_{s_{t,i}}  \Biggl(& w_{f_j^\ast(\mathbf{x_t},\mathbf{s_t})}E \left( d_{f_j^\ast(\mathbf{x_t},\mathbf{s_t})}(x_{t,i})\right)- E\left(w_{c_{t,j}}d_{c_{t,j}}(x_{t,i},t)\right)\Biggr),
\end{align}
where $d_{c_{t,j}}(x_{t,i},t)$ denotes the random demand for the cached file $c_{t,j}\in C_t$ of user $i$ with context vector $x_{t,i}$ at time $t$. 
Here, the expectation is taken with respect to the choices made by the learning algorithm and the distributions of the demands.

\section{A Context-Aware Proactive Caching Algorithm}\label{Sec_ContextAlgo}
In order to proactively cache the most suitable files given the context information about currently connected users, the caching entity should learn context-specific content popularity.
Due to the above formal problem formulation, this problem corresponds to a contextual multi-armed bandit problem and we can adapt and extend a contextual learning algorithm~\cite{Tekin.Schaar2015, TekinZhangSchaar2014} to our setting.
Our algorithm is based on the assumption that users with similar context information will request similar files. 
If this natural assumption holds true, the users' context information together with their requests for cached files can be exploited to learn for future caching decisions.
For this purpose, our algorithm starts by partitioning the context space uniformly into smaller sets, i.e., it splits the context space into parts of similar contexts. 
Then, the caching entity learns the content popularity independently in each of these sets of similar contexts. 
The algorithm operates in discrete time slots. 
In each time slot, the algorithm first observes the contexts of currently connected users. 
Then, the algorithm selects which files to cache in this time slot. 
Based on a certain control function, the algorithm is either in an exploration phase, in which it chooses a random set of files to cache. 
Theses phases are needed to learn the popularity of files which have not been cached often before. 
Otherwise, the algorithm is in an exploitation phase, in which it caches files which on average were requested most when cached in previous time slots with similar user contexts. 
After caching the new set of files, the algorithm observes the users' requests for these files. 
In this way, over time, the algorithm learns context-specific content popularity.  

\begin{figure}
\textbf{m-CAC: Context-Aware Proactive Caching Algorithm}
\begin{algorithmic}[1]
\State Input:  $T$, $h_T$, $K(t)$
\State Initialize context partition: Create partition $\mathcal{P}_T$ of context space $[0,1]^D$ into $(h_T)^D$ hypercubes of identical size 
\State Initialize counters: For all $f\in F$ and all $p \in \mathcal{P}_T$, set
$N_{f,p}=0$ 
\State Initialize estimated demands: For all $f\in F$ and all $p \in \mathcal{P}_T$, set
$\hat{d}_{f,p}=0$
\For{\textbf{each} $t=1,...,T$}
	\State Observe number $U_t$ of currently connected users
	\State Observe user contexts $\mathbf{x_t}=(x_{t,i})_{i=1,...,U_t}$ and service types $\mathbf{s_t}=(s_{t,i})_{i=1,...,U_t}$
	\State Find $\mathbf{p_t}=(p_{t,i})_{i=1,...,U_t}$ such that $x_{t,i} \in p_{t,i}\in \mathcal{P}_T, i=1,...,U_t$
	\State Compute the set of under-explored files $F_{\mathbf{p_t}}^{\ue}(t)$ in~\eqref{Set_underexplored}
	\If{$F_{\mathbf{p_t}}^{\ue}(t)\neq \emptyset$}\Comment{Exploration}
		\State $u = \size(F_{\mathbf{p_t}}^{\ue}(t))$
			\If{$u\geq m$}
				\State Select $c_{t,1},...,c_{t,m}$ randomly from $F_{\mathbf{p_t}}^{\ue}(t)$
			\Else
				\State Select  $c_{t,1},...,c_{t,u}$ as the $u$ files from $F_{\mathbf{p_t}}^{\ue}(t)$
				\State Select  $c_{t,u+1},...,c_{t, m}$ as the $(m-u)$ files $\hat{f}_{1,\mathbf{p_t},\mathbf{s_t}}(t),...,\hat{f}_{m-u,\mathbf{p_t},\mathbf{s_t}}(t)$ from \eqref{Files_select1}
			\EndIf
	\Else \Comment{Exploitation}
		\State Select  $c_{t,1},...,c_{t,m}$ as the $m$ files $\hat{f}_{1,\mathbf{p_t},\mathbf{s_t}}(t),...,\hat{f}_{m,\mathbf{p_t},\mathbf{s_t}}(t)$ from \eqref{Files_select2}
	\EndIf
	\State Observe demand $(d_{j,i})$ of each user $i=1,...,U_t$ for each file $c_{t,j}, j=1,...,m$
	\For{$i=1,...,U_t$}
	\For{$j=1,...,m$}
	\State  $\hat{d}_{c_{t,j},p_{t,i}} = \frac{\hat{d}_{c_{t,j},p_{t,i}}N_{c_{t,j},p_{t,i}} + d_{j,i}}{N_{c_{t,j},p_{t,i}} + 1}$ and $N_{c_{t,j},p_{t,i}} = N_{c_{t,j},p_{t,i}} + 1$
	\EndFor
	\EndFor
\EndFor
\end{algorithmic}
\caption{Pseudocode of m-CAC.}
\label{Algo_uniform}
\end{figure}

The algorithm for selecting $m$ files is called \textit{Context-Aware Proactive Caching with Cache Size~m} (m-CAC) and its pseudocode is given in Fig.~\ref{Algo_uniform}.
Next, we describe the algorithm in more detail. 
In its initialization phase, m-CAC creates a partition $\mathcal{P}_T$ of the context space $\mathcal{X}=[0,1]^D$ into $(h_T)^D$ sets, that are given by $D$-dimensional hypercubes of identical size $\frac{1}{h_T}\times \hdots \times\frac{1}{h_T}$. 
Here, $h_T$ is an input parameter which determines the number of sets in the partition. 
Additionally, m-CAC keeps a counter $N_{f,p}(t)$ for each pair consisting of a file $f \in F$ and a set $p \in \mathcal{P}_T$. 
The counter $N_{f,p}(t)$ is the number of times in which file $f \in F$ was cached after a user with context from set $p$ was connected to the caching entity up to time slot~$t$ (i.e., if 2 users with context from set $p$ are connected in one time slot and file $f$ is cached, this counter is increased by 2). 
Moreover, m-CAC keeps the estimated demand $\hat{d}_{f,p}(t)$ up to time slot $t$ of each pair consisting of a file $f \in F$ and a set $p\in\mathcal{P}_T$. 
This estimated demand is calculated as follows: Let $\mathcal{E}_{f,p}(t)$ be the set of observed demands of users with context from set $p$ when file $f$ was cached up to time slot $t$. 
Then, the estimated demand of file $f$ in set $p$ is given by the sample mean $\hat{d}_{f,p}(t):=\frac{1}{|\mathcal{E}_{f,p}(t)|} \sum_{d\in \mathcal{E}_{f,p}(t)} d$.\footnote{The set $\mathcal{E}_{f,p}(t)$ does not have to be stored since the estimated demand $\hat{d}_{f,p}(t)$ can be updated based on $\hat{d}_{f,p}(t-1)$, $N_{f,p}(t-1)$ and on the observed demands at time $t$.}$^{,}$\footnote{Note that in the pseudocode in Fig.~\ref{Algo_uniform}, the argument $t$ is dropped from counters $N_{f,p}(t)$ and $\hat{d}_{f,p}(t)$ since previous values of these counters do not have to be stored.}

In each time slot $t$, m-CAC first observes the number of currently connected users $U_t$, their contexts $\mathbf{x_t}=(x_{t,i})_{i=1,...,U_t}$ and the service types 
$\mathbf{s_t} = (s_{t,i})_{i=1,...,U_t}$. 
For each context vector $x_{t,i}$, m-CAC determines the set $p_{t,i} \in \mathcal{P}_T$, to which the context vector belongs, i.e., such that $x_{t,i} \in p_{t,i}$ holds.
The collection of these sets is given by $\mathbf{p_t}=(p_{t,i})_{i=1,...,U_t}$. 
Then, the algorithm can either be in an exploration phase or in an exploitation phase. 
In order to determine the correct phase in the current time slot, the algorithm checks if there are files that have not been explored sufficiently often. 
For this purpose, the \textit{set of under-explored files} $F_{\mathbf{p_t}}^{\ue}(t)$ is calculated based on
\begin{align}\label{Set_underexplored}
F_{\mathbf{p_t}}^{\ue}(t)&:=\cup_{i=1}^{U_t} F_{p_{t,i}}^{\ue}(t)\notag\\
&:=\cup_{i=1}^{U_t} \{f\in F: N_{f,p_{t,i}}(t)\leq K(t)\},
\end{align}
where $K(t)$ is a deterministic, monotonically increasing control function, which is an input to the algorithm. 
The control function has to be set adequately to balance the trade-off between exploration and exploitation. 
In Section~\ref{Sec_regret}, we will select a control function that guarantees a good balance in terms of this trade-off. 

If the set of under-explored files is non-empty, m-CAC enters the exploration phase. 
Let $u(t)$ be the size of the set of under-explored files. 
If the set of under-explored files contains at least $m$ elements, i.e., $u(t)\geq m$, the algorithm randomly selects $m$ files from $F_{\mathbf{p_t}}^{\ue}(t)$ to cache. 
If the set of under-explored files contains less than $m$ elements, i.e., $u(t)<m$, it selects all $u(t)$ files from $F_{\mathbf{p_t}}^{\ue}(t)$ to cache. 
Since the cache is not fully filled by $u(t)<m$ files, $(m-u(t))$ additional files can be cached. 
In order to exploit knowledge obtained so far, m-CAC selects $(m-u(t))$ files from $F\setminus F_{\mathbf{p_t}}^{\ue}(t)$ based on a file ranking according to the estimated weighted demands, 
as defined by the files $\hat{f}_{1,\mathbf{p_t},\mathbf{s_t}}(t),...,\hat{f}_{m-u(t),\mathbf{p_t},\mathbf{s_t}} (t)\in F\setminus F_{\mathbf{p_t}}^{\ue}(t)$, which satisfy for $j=1,..., m-u(t)$:
\begin{align}\label{Files_select1}
\hat{f}_{j,\mathbf{p_t},\mathbf{s_t}}(t)&\in \argmax _{f\in F\setminus (F_{\mathbf{p_t}}^{\ue}(t) \cup \bigcup\limits_{k=1}^{j-1}\{\hat{f}_{k,\mathbf{p_t},\mathbf{s_t}}(t)\})}  w_f\sum_{i=1}^{U_t}v_{s_{t,i}}\hat{d}_{f,p_{t,i}}(t).
\end{align}
If the set of files defined by~\eqref{Files_select1} is not unique, ties are broken arbitrarily.
Note that by this procedure, even in exploration phases, the algorithm additionally exploits, whenever the number of under-explored files is smaller than the cache size.

If the set of under-explored files $F_{\mathbf{p_t}}^{\ue}(t)$ is empty, m-CAC enters the exploitation phase. 
It selects $m$ files from $F$ based on a file ranking according to the estimated weighted demands, as defined by the files $\hat{f}_{1,\mathbf{p_t},\mathbf{s_t}}(t),...,\hat{f}_{m,\mathbf{p_t},\mathbf{s_t}} (t)\in F$, which satisfy for $j=1,..., m$:
\begin{align}\label{Files_select2}
\hat{f}_{j,\mathbf{p_t},\mathbf{s_t}}(t)&\in \argmax _{f\in F\setminus \left(\bigcup_{k=1}^{j-1}\{\hat{f}_{k,\mathbf{p_t},\mathbf{s_t}}(t)\}\right)}  w_f\sum_{i=1}^{U_t}v_{s_{t,i}}\hat{d}_{f,p_{t,i}}(t).
\end{align}
If the set of files defined by \eqref{Files_select2} is not unique, ties are again broken arbitrarily.

After selecting the subset of files to cache, the algorithm observes the users' requests for these files in this time slot. 
Then, it updates the estimated demands and the counters of cached files.

\section{Analysis of the Regret}\label{Sec_regret}
In this section, we give an upper bound on the regret $R(T)$ of m-CAC in~\eqref{Eq_regret}.
The regret bound is based on the natural assumption that expected demands for files are similar in similar contexts, i.e., that users with similar characteristics are likely to consume similar content. This assumption is realistic since the users' preferences in terms of consumed content differ based on the users' contexts, so that it is plausible to divide the user population into segments of users with similar context and similar preferences. 
Formally, the similarity assumption is captured by the following H\"older condition.

\begin{Assumption} \label{Ass_Hoelder}
There exists $L>0$, $\alpha>0$ such that for all $f\in F$ and for all $x,y \in \mathcal{X}$, it holds that
\begin{align*}
|\mu_f(x)-\mu_f(y)|\leq L ||x-y||^{\alpha},
\end{align*}
where $||\cdot||$ denotes the Euclidean norm in $\mathbb{R}^D$. 
\end{Assumption}

Assumption~\ref{Ass_Hoelder} is needed for the analysis of the regret, but it should be noted that m-CAC can also be applied if this assumption does not hold true. 
However, a regret bound might not be guaranteed in this case.

The following theorem shows that the regret of m-CAC is sublinear in the time horizon $T$, i.e., $R(T) = O(T^{\gamma})$ with $\gamma<1$. 
This bound on the regret guarantees that the algorithm has an asymptotically optimal performance, since then $\lim _{T\rightarrow \infty} \frac{R(T)}{T} = 0$ holds. 
This means, that m-CAC converges to the oracle solution strategy. 
In other words, m-CAC converges to the optimal cache content placement strategy, which maximizes the expected number of cache hits. 
In detail, the regret of m-CAC can be bounded as follows for any finite time horizon $T$.

\begin{Theorem}[Bound for $R(T)$]\label{Theorem_Regret}
Let $K(t) = t^{\frac{2\alpha}{3 \alpha + D}} \log (t)$ and $h_T = \ceil{T^{\frac{1}{3\alpha + D}}}$. If m-CAC is run with these parameters and Assumption \ref{Ass_Hoelder} holds true, 
the leading order of the regret $R(T)$ is $O\left(v_{\max} w_{\max} m U_{\max}R_{\max} |F| T^{\frac{2\alpha + D}{3\alpha + D}} \log(T) \right)$.
\end{Theorem}

The proof can be found in our online appendix \cite{appendix}.
The regret bound given in Theorem~\ref{Theorem_Regret} is sublinear in the time horizon $T$,
proving that m-CAC converges to the optimal cache content placement strategy. 
Additionally, Theorem~\ref{Theorem_Regret} is applicable for any finite time horizon $T$, such that it provides a bound on the loss incurred by m-CAC for any finite number of cache placement phases. 
Thus, Theorem~\ref{Theorem_Regret} characterizes \mbox{m-CAC's} speed of convergence
Furthermore, Theorem~\ref{Theorem_Regret} shows that the regret bound is a constant multiple of the regret bound in the special case without service differentiation, in which $v_{\max}=  1 $ and $w_{\max}=1$. 
Hence, the order of the regret is $O\left(T^{\frac{2\alpha + D}{3\alpha + D}} \log(T)\right)$ in the special case as well.

\section{Memory Requirements}\label{Sec_complexity}
The memory requirements of m-CAC are mainly determined by the counters kept by the algorithm during its runtime (see also~\cite{Tekin.Schaar2015}).
For each set $p$ in the partition $\mathcal{P}_T$ and each file $f\in F$, the algorithm keeps the counters $N_{f,p}$ and $\hat{d}_{f,p}$. 
The number of files is $|F|$. 
If m-CAC runs with the parameters from Theorem~\ref{Theorem_Regret}, the number of sets in $\mathcal{P}_T$ is upper bounded by $(h_T)^D = \ceil{T^{\frac{1}{3\alpha + D}}}^D \leq 2^D T^{\frac{D}{3\alpha + D}}$. 
Hence, the required memory is upper bounded by $|F| 2^D T^{\frac{D}{3\alpha + D}}$ and is thus sublinear in the time horizon $T$.
This means, that for $T\rightarrow \infty$, the algorithm would require infinite memory.
However, for practical approaches, only the counters of such sets $p$ have to be kept 
to which at least one of the connected users' context vectors has already belonged to.
Hence, depending on the heterogeneity in the connected users' context vectors, the required number of counters that have to be kept can be much smaller than given by the upper bound.

\section{Extensions}\label{Sec_Extensions}

\subsection{Exploiting the Multicast Gain}
So far, we assumed that each request for a cached file is immediately served by a unicast transmission.
However, our algorithm can be extended to multicasting, which has been shown to be beneficial in combination with caching~\cite{Maddah-AliNiesen2014, PoularakisIosifidisSourlasEtAl2016}.
For this purpose, to extend our algorithm, each time slot $t$ is divided into a fixed number of intervals. 
In each interval, incoming requests are monitored and accumulated. 
At the end of the interval, requests for the same file are served by a multicast transmission.
In order to exploit knowledge about content popularity learned so far, a request for a file with low estimated demand could, however, still be served by a unicast transmission. 
In this way, unnecessary delays are prevented in cases in which another request and thus a multicast transmission are not expected. 
Moreover, service differentiation could be taken into account. 
For example, high-priority users could be served by unicast transmissions, such that their delay is not increased due to waiting times for multicast transmissions.

\subsection{Rating-Based Context-Aware Proactive Caching}
So far, we considered cache content placement with respect to the demands $d_f(x)$ in order
to maximize the number of (weighted) cache hits.
However, a CP operating an infostation might want to cache not only content that is requested often, but which also receives high ratings from the users. 
Consider the case that after consumption users rate content in a range $[r_{\min},r_{\max}]\subset \mathbb{R}_+$. 
For a context $x$, let $r_f(x)$ be the random variable describing the rating of a user with context $x$ if he requests file $f$ and makes a rating thereafter.
Then, we define the random variable
\begin{align}\label{Def_rating_based_caching}
\tilde{d}_f(x) &:= r_f(x) d_f(x),
\end{align}
which combines the demand and the rating for file $f$ of a user with context $x$. 
By carefully designing the range of ratings, the CP chooses the trade-off between ratings and cache hits. 
Now, we can apply m-CAC with respect to $\tilde{d}_f(x)$. 
In this case, m-CAC additionally needs to observe the users' ratings in order to learn content popularity in terms of ratings. 
If the users' ratings are always available, Theorem~\ref{Theorem_Regret} applies and provides a regret bound of $O\left(v_{\max} w_{\max} r_{\max} m U_{\max}R_{\max} |F| T^{\frac{2\alpha + D}{3\alpha + D}} \log(T) \right)$.

However, users might not always reveal a rating after consuming a content. 
When a user's rating is missing, we assume that m-CAC does not update the counters based on this user's request. 
This may result in a higher required number of exploration phases. 
Hence, the regret of the learning algorithm is influenced by the users' willingness to reveal ratings of requested content.
Let $q\in (0,1)$ be the probability that a user reveals his rating after requesting a file.
Then, the regret of the learning algorithm is bounded as given below.

\begin{Theorem}[Bound for $R(T)$ for rating-based caching with missing ratings]\label{Theorem_Regret_Rating}
Let $K(t) = t^{\frac{2\alpha}{3 \alpha + D}} \log (t)$ and $h_T = \ceil{T^{\frac{1}{3\alpha + D}}}$. 
If m-CAC is run with these parameters with respect to $\tilde{d}_f(x)$, Assumption \ref{Ass_Hoelder} holds true for $\tilde{d}_f(x)$ and a user reveals his rating with probability $q$, the leading order of the regret $R(T)$ is $O\left(\frac{1}{q} v_{\max} w_{\max} r_{\max} m U_{\max}R_{\max} |F| T^{\frac{2\alpha + D}{3\alpha + D}} \log(T) \right)$.
\end{Theorem}
The proof can be found in our online appendix~\cite{appendix}.
Comparing Theorem~\ref{Theorem_Regret_Rating} with Theorem \ref{Theorem_Regret}, the regret of m-CAC is scaled up by a factor $\frac{1}{q}>1$ in case of rating-based caching with missing ratings. 
This factor corresponds to the expected number of requests until the caching entity receives one rating. 
However, the time order of the regret remains the same.
 Hence, m-CAC is robust under missing ratings in the sense that if some users refuse to rate requested content, the algorithm still converges to the optimal cache content placement strategy.

\subsection{Asynchronous User Arrival}
So far, we assumed that the set of currently connected users only changes in between the time slots of our algorithm. 
This means, that only those users connected to the caching entity at the beginning of a time slot, 
will request files within that time slot.
However, if users connect to the caching entity asynchronously, m-CAC should be adapted.
If a user directly disconnects after the context monitoring without requesting any file, he should be excluded from learning. 
Hence, in m-CAC, the counters are not updated for disconnecting users.
If a user connects to the caching entity after cache content placement, his context was not considered in the caching decision. 
However, his requests can be used to learn faster.
Hence, in m-CAC, the counters are updated based on this user's requests.

\subsection{Multiple Wireless Local Caching Entities}

So far, we considered online learning for cache content placement in a single caching entity. 
However, real caching systems contain multiple caching entities, each of which should learn local content popularity. 
In a network of multiple caching entities, m-CAC could be applied separately and independently by each caching entity. 
For the case that coverage areas of caching entities overlap, in this subsection, we present m-CACao, an extension of m-CAC to \textit{Context-Aware Proactive Caching with Area Overlap}. 
The idea of m-CACao is that caching entities can learn content popularity faster by not only relying on their own cache hits, but also on cache hits at neighboring caching entities with overlapping coverage area. 
For this purpose, the caching entities overhear cache hits produced by users in the intersection to neighboring coverage areas.

In detail, m-CAC is extended to m-CACao as follows:
The context monitoring and the selection of cache content works as in m-CAC. 
However, the caching entity not only observes its own cache hits (line 21 in Fig.~\ref{Algo_uniform}), but it overhears cache hits at neighboring caching entities of users in the intersection. 
Then, the caching entity not only updates the counters of its own cached files (lines 22-26 in Fig.~\ref{Algo_uniform}), but it additionally updates the counters of files of which it overheard cache hits at neighboring caches.
This helps the caching entity to learn faster.

\section{Numerical Results} \label{Sec_NumResults}
In this section, we numerically evaluate the proposed learning algorithm m-CAC 
by comparing its performance to several reference algorithms based on a real world data set.

\subsection{Description of the Data Set}
We use a data set from MovieLens~\cite{HarperKonstan2015} to evaluate our proposed algorithm.
MovieLens is an online movie recommender operated by the research group GroupLens from the University of Minnesota.
The MovieLens 1M DataSet~\cite{movielens} contains 1000209 ratings of 3952 movies.
These ratings were made by 6040 users of MovieLens within the years 2000 to 2003. 
Each data set entry consists of an anonymous user ID, a movie ID, a rating (in whole numbers between 1 and 5) and a timestamp. 
Additionally, demographic information about the users is given: Their gender, age (in 7 categories), occupation (in 20 categories) as well as their Zip-code.
For our numerical evaluations, we assume that the movie rating process in the data set corresponds to a content request process of users connected to a wireless local caching entity (see~\cite{Li2016},~\cite{LiXuSchaarEtAl2016a} for a similar approach). 
Hence, a user rating a movie at a certain time in the data set for us corresponds to a request to either the caching entity (in case the movie is cached in the caching entity) or to the macro cellular network (in case the movie is not cached in the caching entity).
This approach is reasonable since users typically rate movies after watching them.

\begin{figure}[!t]
\centering
\includegraphics[width=0.6\textwidth]{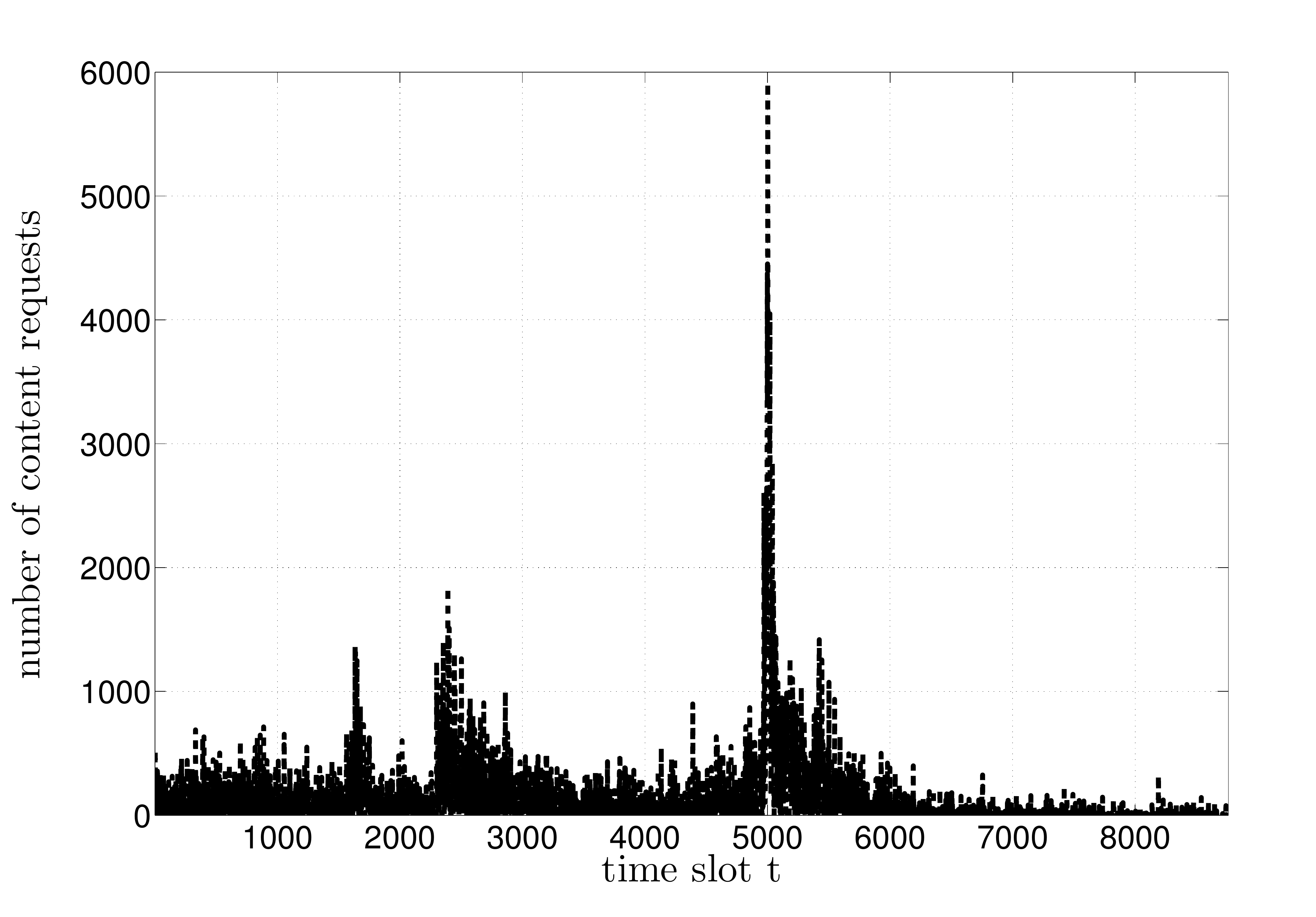}
\caption{Number of content requests in used data set as a function of time slots. Time slots at an hourly basis.}
\label{Fig_content_requests}
\end{figure}

In our simulations, we only use the data gathered within the first year of the data set,
since around $94 \%$ of the ratings were provided within this time frame.
Then, we divide a year's time into $8760$ time slots of one hour each ($T=8760$), assuming that the caching entity updates its cache content at an hourly basis. Then, we assign the requests and corresponding user contexts to the time slots according to their timestamps and we interpret each request as if it was coming from a separate user. 
At the beginning of a time slot, we assume to have access to the context of each user responsible for a request in the coming time slot. 
Fig.~\ref{Fig_content_requests} shows that the corresponding content request process is bursty and flattens out towards the end.
As context dimensions, we select the dimensions gender and age.\footnote{We neglect the occupation as context dimension since by mapping them to a [0,1] variable, we would have to classify which occupations are more and which are less similar to each other.}

\subsection{Reference Algorithms}
We compare m-CAC with five reference algorithms. 
The first algorithm is the omniscient Oracle, which has complete knowledge about the exact future demands.
In each time slot, the oracle selects the optimal $m$ files that will maximize the number of cache hits in this time slot.\footnote{Note that this oracle yields even better results than the oracle used as a benchmark to define the regret in~\eqref{Eq_regret}. 
In the definition of regret, the oracle only exploits knowledge about expected demands, instead of exact future demands.}

The second reference algorithm is called m-UCB, which consists of a variant of the UCB algorithm.
UCB is a classical learning algorithm for multi-armed bandit problems~\cite{Auer.etal2002},
which has logarithmic regret order.
However, it does not take into account context information, 
i.e., the logarithmic regret is with respect to the average expected demand over the whole context space.
While in classical UCB, one action is taken in each time slot,
we modify UCB to take $m$ actions at a time, which corresponds to selecting $m$ files.

The third reference algorithm is the m-$\epsilon$-Greedy.
This is a variant of the simple $\epsilon$-Greedy~\cite{Auer.etal2002} algorithm, 
which does not consider context information. 
The m-$\epsilon$-Greedy caches a random set of $m$ files with probability $\epsilon \in (0,1)$.
With probability $(1-\epsilon)$, the algorithm caches the $m$ files with highest to
$m$-th highest estimated demands.
These estimated demands are calculated based on previous demands for cached files.

The fourth reference algorithm is called m-Myopic. 
This is an algorithm taken from~\cite{Blasco.Gunduz2014a}, which is investigated since it is 
comparable to the well-known Least Recently Used algorithm (LRU) for caching.
m-Myopic only learns from one time slot in the past. 
It starts with a random set of files and in each of the following time slots discards all files that have not been requested in the previous time slot.
Then, it randomly replaces the discarded files by other files. 

The fifth reference algorithm, called Random, caches a random set of files in each time slot.

\subsection{Performance Measures}
The following performance measures are used in our analysis.
The evolution of per-time slot or cumulative \textit{number of cache hits} allows comparing the absolute performance of the algorithms.
A relative performance measure is given by the \textit{cache efficiency}, 
which is defined as the ratio of cache hits compared to the overall demand, i.e.,
\begin{align*}
\text{cache efficiency in }\%=\frac{\text{cache hits}}{\text{cache hits} + \text{cache misses}}\cdot 100.
\end{align*}
The cache efficiency describes the percentage of requests which can be served by cached files. 

\subsection{Results}
\begin{figure}[!t]
\centering
\subfigure[Number of cache hits per time slot.]{\label{Fig_sim_m200_fix_single}\includegraphics[width=0.6\textwidth]{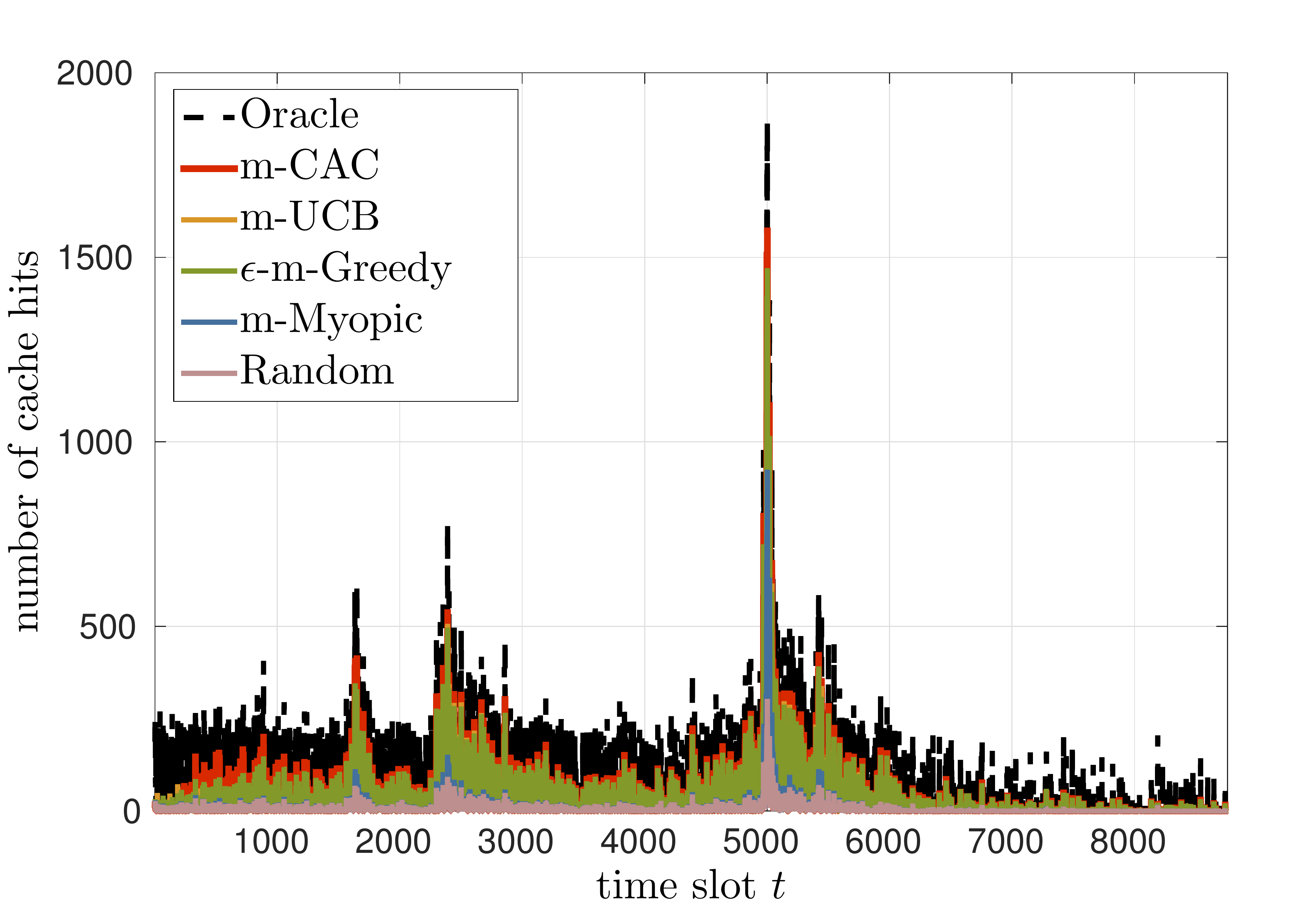}}
\hfil
\subfigure[Cumulative number of cache hits.]{\label{Fig_sim_m200_fix_aggr}\includegraphics[width=0.6\textwidth]{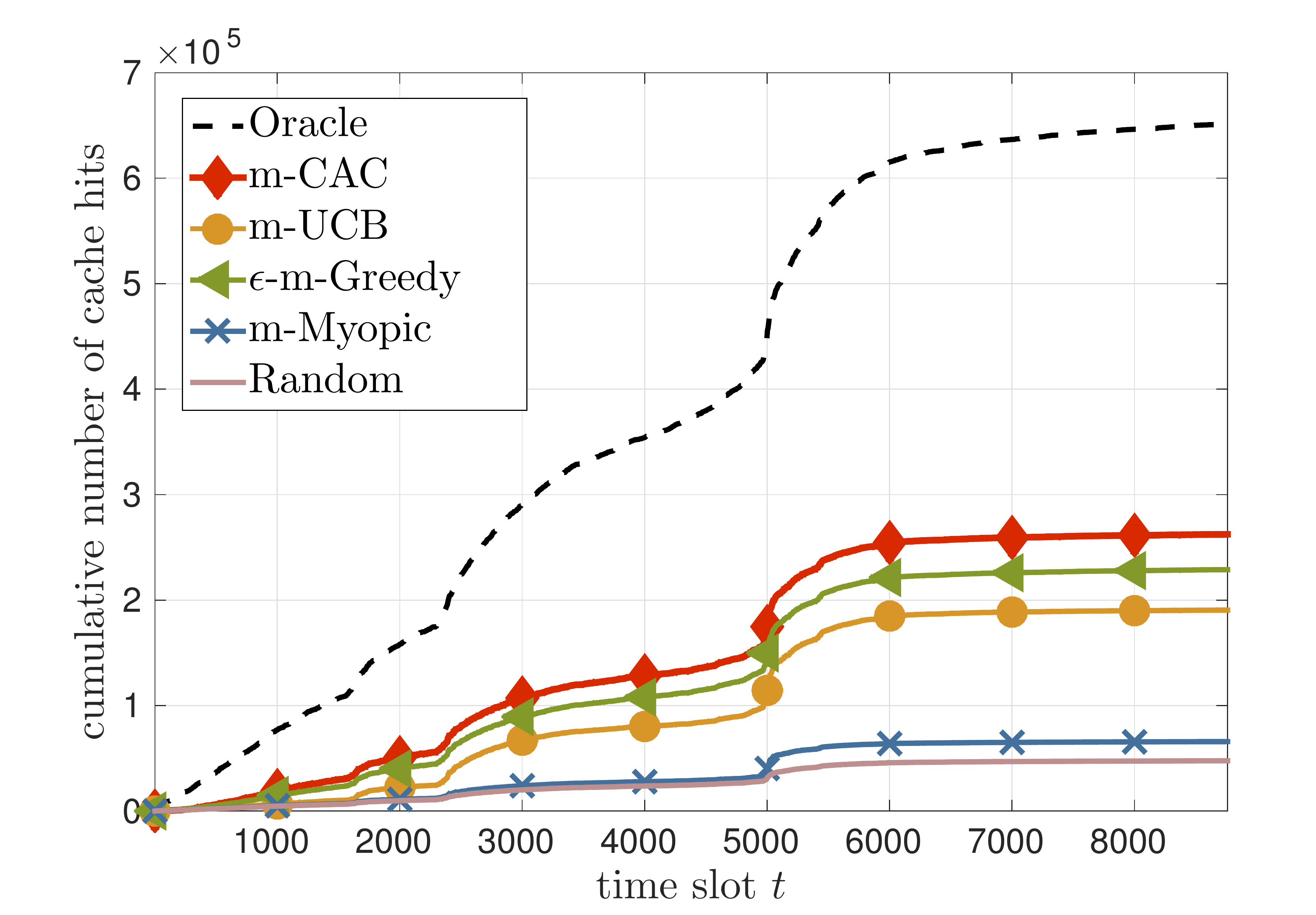}}
\caption{Time evolution of algorithms for $m=200$.}
\label{Fig_sim_m200}
\end{figure}

In our simulations, we set $\epsilon = 0.09$ in m-$\epsilon$-Greedy, which is the value at which heuristically the algorithm on average performed best.
In m-CAC, we set the control function to $K(t) = c \cdot t^{\frac{2\alpha}{3 \alpha + D}} \log (t)$
with $c=1/(|F|D)$.\footnote{Compared to the control function in Theorem~\ref{Theorem_Regret}, the additional factor reduces the number of exploration phases which allows for better performance.}
The simulation results are obtained by averaging over $100$ runs of the algorithms.
First, we consider the case without service differentiation.
The long-term behavior of m-CAC is investigated with the following scenario. 
We assume that the caching entity can store $m=200$ movies out of the $|F|=3952$ available movies. 
Hence, the cache size corresponds to about $5\%$ of the file library size.
We run all algorithms on the data set and study their results as a function of time, i.e., over the time slots $t=1,...,T$.
Fig.~\ref{Fig_sim_m200_fix_single}~and~\ref{Fig_sim_m200_fix_aggr} show the per-time slot and the cumulative numbers of cache hits up to time slot $t$ as a function of time, respectively. 
Due to the bursty content request process (compare Fig.~\ref{Fig_content_requests}),
also the number of cache hits achieved by the different algorithms is bursty over time.
As expected, the Oracle gives an upper bound to the other algorithms. 
Among the other algorithms, m-CAC, m-$\epsilon$-Greedy and m-UCB clearly outperform m-Myopic and Random. 
This is due to the fact that these three algorithms learn from the history of observed demands, while m-Myopic only learns from one time slot in the past and Random does not learn at all. 
It can be observed that m-$\epsilon$-Greedy shows a better performance than m-UCB, even though it uses a simpler learning strategy. 
Overall, m-CAC outperforms the other algorithms by additionally learning from context information. 
At the time horizon, the cumulative number of cache hits achieved by m-CAC is $1.146$, $1.377$, $3.985$ and $5.506$ times the cumulative number of cache hits achieved by m-$\epsilon$-Greedy, m-UCB, m-Myopic and Random, respectively.

\begin{figure}[!t]
\centering
\includegraphics[width=0.6\textwidth]{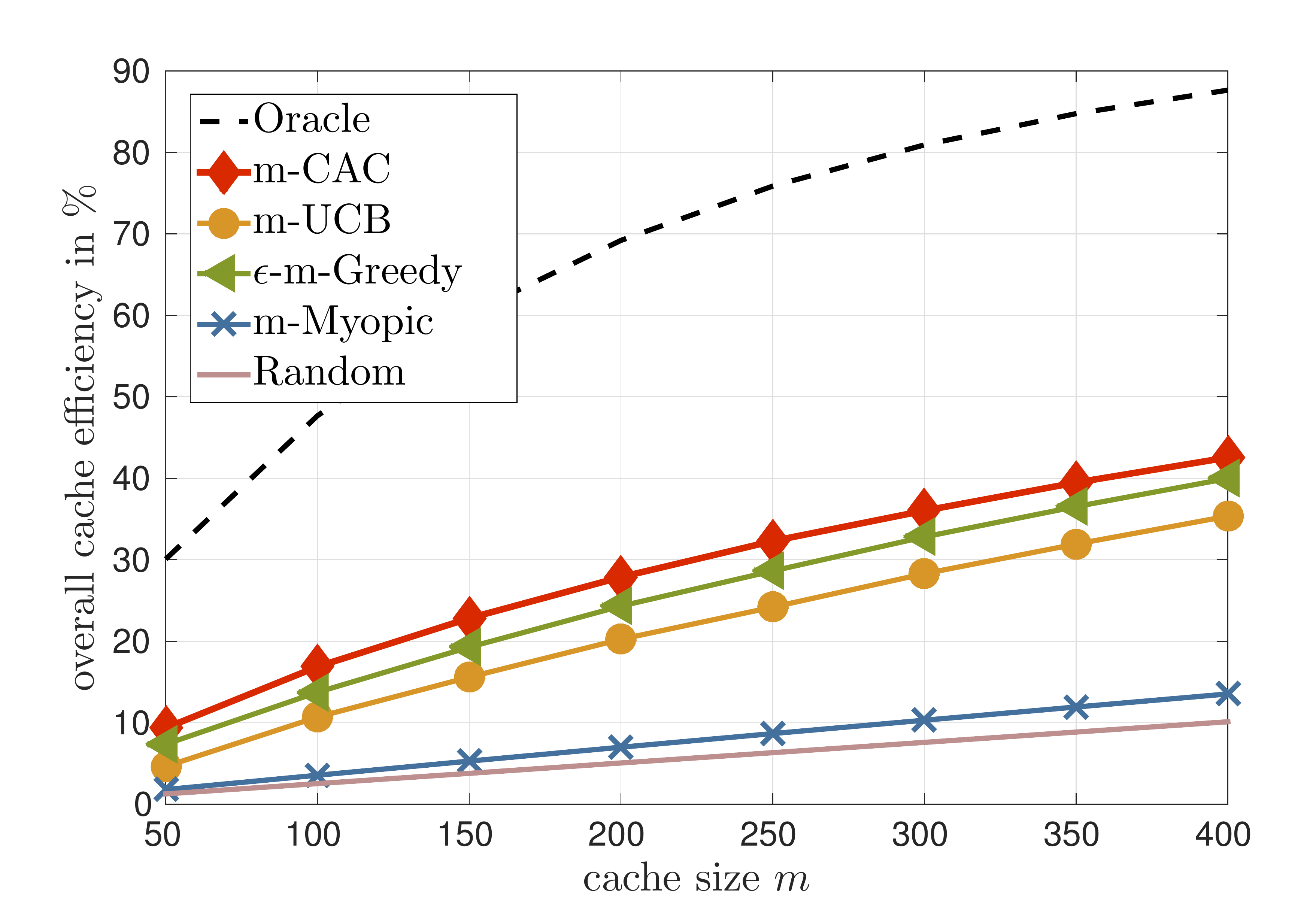}
\caption{Overall cache efficiency at $T$ as a function of cache size $m$.}
\label{Fig_sim_m_all_fix_aggr_eff}
\end{figure}

Next, we investigate the impact of the cache size $m$ by varying it between $50$ and $400$ files, which corresponds to between $1.3\%$ and $10.1\%$ of the file library size,
which is a realistic assumption.
All remaining parameters are kept as before.
Fig.~\ref{Fig_sim_m_all_fix_aggr_eff} shows the overall cache efficiency achieved at the time horizon $T$ as a function of cache size, i.e., the cumulative number of cache hits up to $T$ is normalized by the cumulative number of requests up to $T$.
The overall cache efficiency of all algorithms is increasing with increasing cache size.
Moreover, the results indicate that again m-CAC and m-$\epsilon$-Greedy slightly outperform m-UCB and clearly outperform m-Myopic and Random.
Averaged over the range of cache sizes, the cache efficiency of m-CAC is $28.4\%$, compared to an average cache efficiency of $25.3\%$, $21.4\%$, $7.76\%$ and $5.69\%$ achieved by m-$\epsilon$-Greedy, m-UCB, m-Myopic and Random, respectively.

\begin{figure}[!t]
\centering
\includegraphics[width=0.6\textwidth]{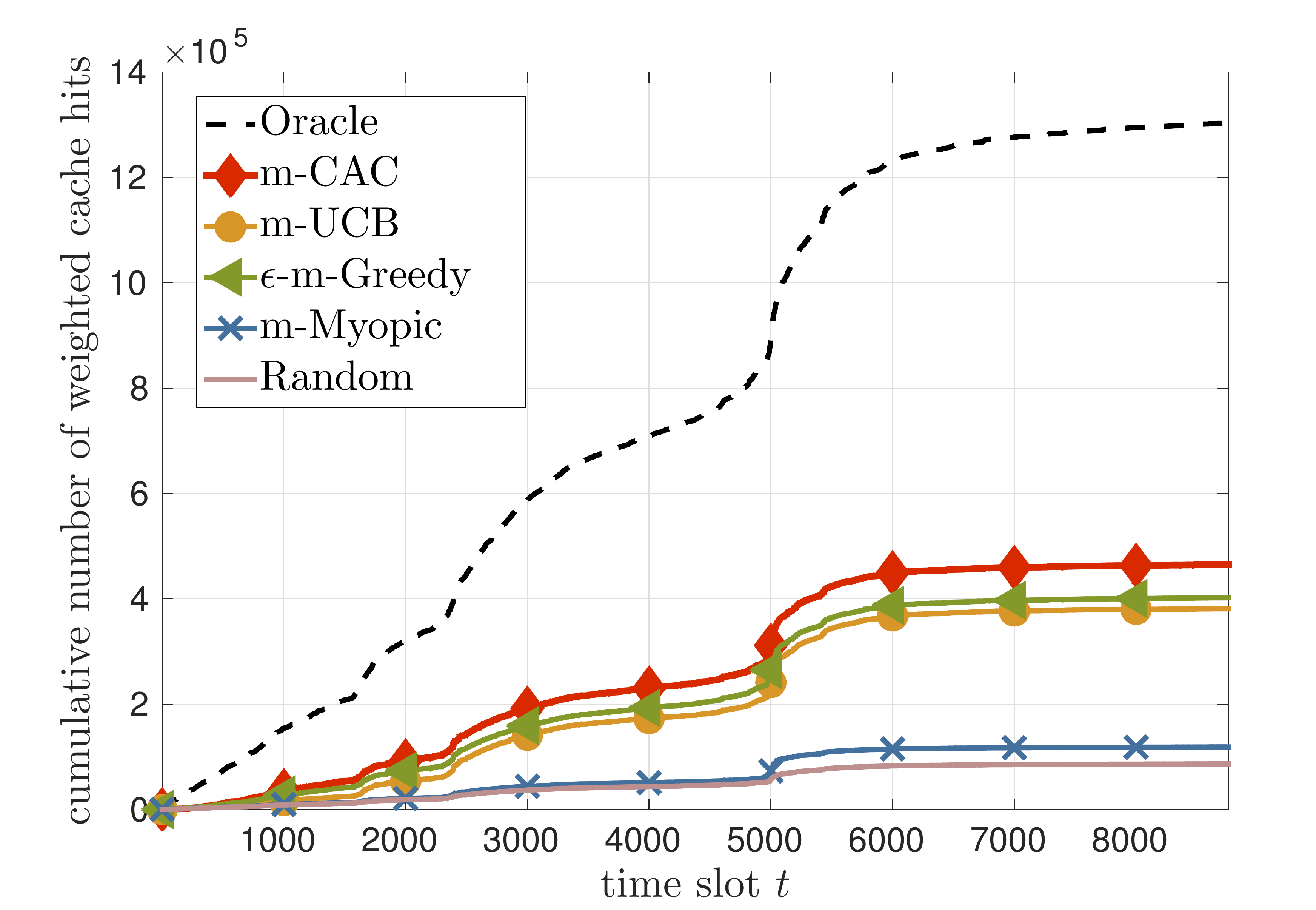}
\caption{Cumulative number of weighted cache hits for $m=200$ as a function of time.}
\label{Fig_sim_m200_fix_aggr_sd01_w5}
\end{figure}

Now, we consider a case of service differentiation, in which two different service types~$1$~and~$2$ with weights $v_{1}= 5$ and $v_{2} = 1$ exist. 
Hence, service type $1$ should be prioritized due to the higher value it represents. 
We randomly assign $10 \%$ of the users to service type $1$ and classify all remaining users as service type $2$. 
Then, we adjust each algorithm to take into account service differentiation by incorporating the weights according to the service types.
Fig.~\ref{Fig_sim_m200_fix_aggr_sd01_w5} shows the cumulative number of weighted cache hits up to time slot $t$ as a function of time. 
At the time horizon, the cumulative number of weighted cache hits achieved by m-CAC is $1.156$, $1.219$, $3.914$ and $5.362$ times the cumulative number of cache hits achieved by m-$\epsilon$-Greedy, m-UCB, m-Myopic and Random, respectively.
A comparison with Fig.~\ref{Fig_sim_m200_fix_aggr} shows that the behavior of the algorithms is similar to the case without service differentiation.

\begin{figure}[!t]
\centering
\includegraphics[width=0.6\textwidth]{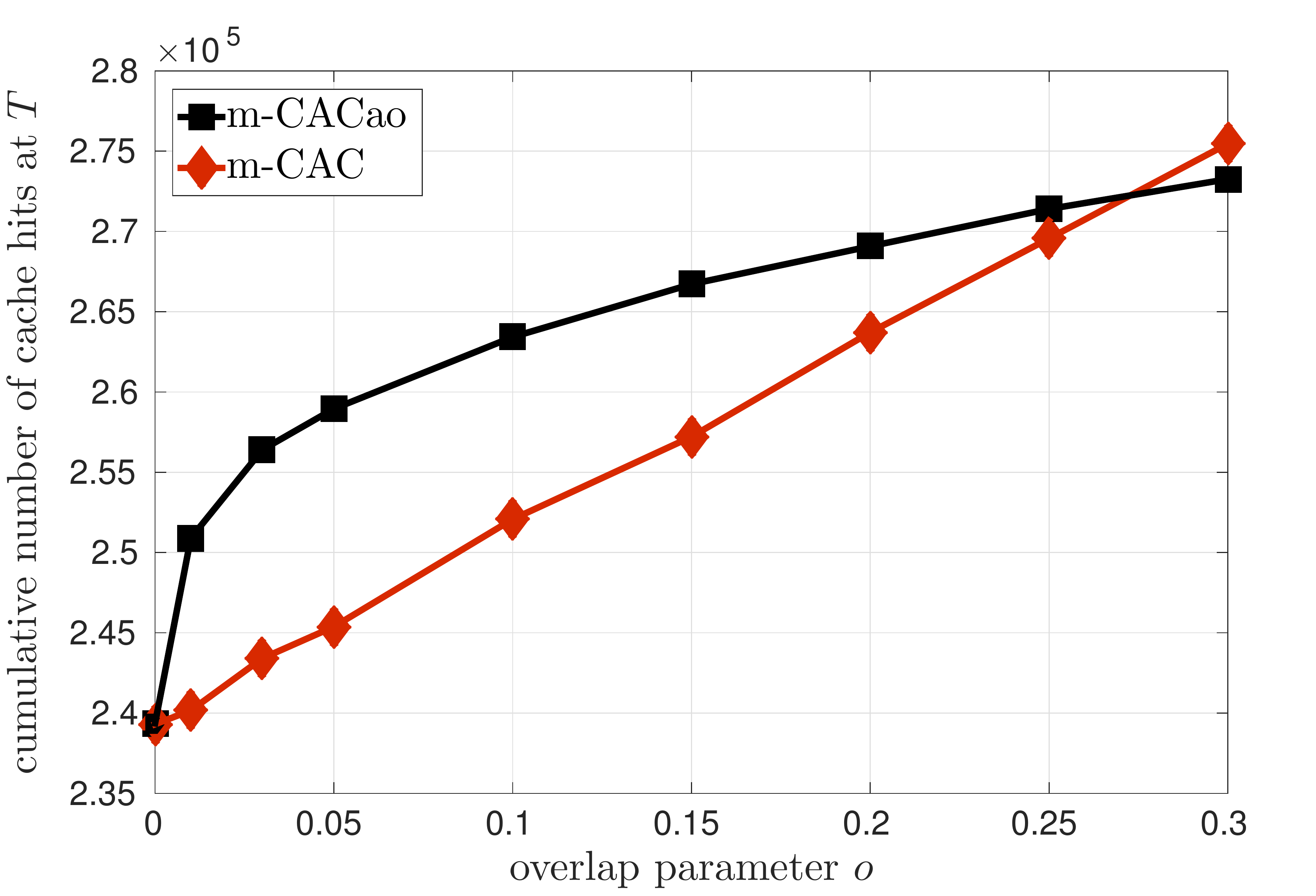}
\caption{Cumulative number of cache hits at $T$ as a function of the overlap parameter $o$.}
\label{Fig_sim_several_caches}
\end{figure}

Finally, we investigate the extension to multiple caching entities and compare the performance of the proposed algorithms m-CAC and m-CACao. 
We consider a scenario with two caching entities and divide the data set as follows: 
A fraction $o\in [0,0.3]$ of randomly selected requests is considered to be made in the intersection of the two coverage areas. 
We use the parameter $o$ as a measure of the overlap between the caching entities. 
The remaining requests are randomly assigned to either one of the caching entities. 
These requests are considered to be made by users solely connected to one caching entity. 
Then, on the one hand we run m-CAC separately on each caching entity and on the other hand we run m-CACao on both caching entities. 
Fig.~\ref{Fig_sim_several_caches} shows the cumulative number of cache hits achieved in sum by the two caching entities at the time horizon $T$ as a function of the overlap parameter $o$. 
As expected, m-CAC and m-CACao perform identically for non-overlapping coverage areas. 
With increasing overlap, the number of cache hits achieved by both m-CAC and m-CACao increases. 
The reason is that users in the intersection can more likely be served since they have access to both caches. 
Hence, even though the caching entities do not coordinate their cache content, more cache hits occur. For up to $25\%$ of overlap ($o\leq 0.25$), m-CACao outperforms m-CAC. 
Clearly, m-CACao performs better since by overhearing cache hits at the neighboring caching entity, both caching entities learn content popularity faster. 
For very large overlap ($o>0.25$), m-CAC yields higher numbers of cache hits. 
The reason is that when applying m-CACao in case of a large overlap, neighboring caching entities overhear such a large number of cache hits, that they learn very similar content popularity distributions. 
Hence, over time it is likely that their caches contain the same files. 
In contrast, applying m-CAC, a higher diversity in cache content is maintained over time. 
Clearly, further gains in cache hits could be achieved by jointly optimizing the cache content of all caching entities. 
However, this would either require coordination among the caching entities or a central planner deciding on the cache content of all caching entities, which results in a high communication overhead. 
In contrast, our heuristic algorithm m-CACao does not require additional coordination or communication and yields good results for small overlaps.

\section{Conclusion}\label{Sec_Conclusion}
In this paper, we presented a context-aware proactive caching algorithm for wireless caching entities based on contextual multi-armed bandits. 
To cope with unknown and fluctuating content popularity among the dynamically arriving and leaving users, the algorithm regularly observes context information of connected users, updates the cache content and subsequently observes cache hits. 
In this way, the algorithm learns context-specific content popularity online, 
which allows for a proactive adaptation of cache content according to fluctuating local content popularity.
We derived a sublinear regret bound, which characterizes the learning speed and proves that our proposed algorithm converges to the optimal cache content placement strategy, which maximizes the expected number of cache hits.
Moreover, the algorithm supports customer prioritization and can be combined with multicast transmissions and rating-based caching decisions.
Numerical studies showed that by exploiting context information, our algorithm outperforms state-of-the-art algorithms in a real world data set.

\section*{Acknowledgment}
The work by S.~M\"uller and A.~Klein has been funded by the German Research Foundation (DFG) as part of projects B3 and C1 within the Collaborative
Research Center (CRC) 1053 -- MAKI. The work by O.~Atan and M.~van~der~Schaar is supported by NSF CCF1524417 and NSF ECCS1407712 grant.


\bibliographystyle{IEEEtran}

\nocite{Hoeffding1963, Chlebus2009}

\end{document}